\documentclass[iop]{emulateapj-rtx4} 
\shortauthors{Sekanina}
\shorttitle{Nebulous Companions to Comet C/1882 R1}
\slugcomment{Version \today }
\newcommand{\Rsun}{$R_{\mbox{\scriptsize \boldmath $\odot$}}\!$}

\begin{document}
\title{ENIGMATIC NEBULOUS COMPANIONS TO THE GREAT SEPTEMBER COMET OF 1882\\
       AS SOHO-LIKE KREUTZ SUNGRAZERS CAUGHT IN TERMINAL OUTBURST}
\author{Zdenek Sekanina}
\affil{Jet Propulsion Laboratory, California Institute of Technology,
  4800 Oak Grove Drive, Pasadena, CA 91109, U.S.A.}
\email{Zdenek.Sekanina@jpl.nasa.gov.}

\begin{abstract} 
I investigate the nature of the transient nebulous companions to the sungrazing comet
C/1882~R1, known as the Great September Comet.  The features were located several degrees
to the southwest of the comet's head and reported independently by four observers, including
J.\ F.\ J.\ Schmidt and E.\ E.\ Barnard, over a period of ten days nearly one month after
perihelion, when the comet was 0.7~AU to 1~AU from the Sun.  I conclude that none of the
nebulous companions was ever sighted more than once and that, contrary to his belief, Schmidt
observed unrelated objects on the four consecutive mornings.  Each nebulous companion is
proposed to have been triggered by a fragment at most a few tens of meters across, released
from the comet's nucleus after perihelion and seen only because it happened to be caught in
the brief terminal outburst, when its mass was suddenly shattered into a cloud of mostly
microscopic debris due possibly to rotational bursting triggered by sublimation torques.  The
fragment's motion was affected by a strong outgassing-driven nongravitational acceleration
with a significant out-of-plane component.  Although fragmentation events were common, only
a small fraction of nebulous companions was detected because of their transient nature.  The
observed brightness of the nebulous companions is proposed to have been due mainly to C$_2$
emissions, with a contribution from scattering of sunlight by the microscopic dust.  By their
nature, the fragments responsible for the nebulous companions bear a strong resemblance to
the dwarf Kreutz sungrazers detected with the coronagraphs aboard the SOHO space probe.  Only
their fragmentation histories are different and the latter display no terminal outburst,
a consequence of extremely short lifetimes of the sublimating icy and refractory material in
the Sun's corona.  
\end{abstract}

\section{Introduction} 
As celestial spectacle, the Great September Comet of 1882 (C/1882~R1) may not have
outclassed its forerunner, the Great March Comet of 1843 (C/1843~D1), another
brilliant Kreutz sungrazer.  Yet, there is no doubt that the 1882 performer attracted
more scientific scrutiny than its rival.  Reasons were many, the major ones being as
follows.

First, at the most general level, technological innovations on the one hand and
scientific progress on the other hand, achieved over the period of four decades
between the arrivals of the two comets meant the availability by the 1880s of
improved astronomical instrumentation operated by a new generation of dedicated,
assiduous, and knowledgeable observers.

Second, cometary science retained its frontline position among the fields of
astronomical research in the late 19th century.  This is illustrated by the
bibliography of E.\ E.\ Barnard, one of the era's most prolific, universal, and
influential observers, who also contributed to the subject of this paper.  A
comprehensive treatise on Barnard's life and work (Frost 1927) shows that the
greatest fraction of his 732~listed publications, 30~percent, did indeed deal
with comets, followed by the planets (23~percent) and nebulae (11~percent).

Third, whereas no bright comet very closely approaching the Sun had been reported
for some 140~years prior to the 1843 sungrazer, one was under observation
over a period of weeks in the early 1880, less than three years before the Great
September Comet.  The 1880 sungrazer provoked a major controversy about whether
it was a return of the 1843 sungrazer (implying the subsequently discredited
orbital period of 37~years) or another fragment of a shared parent comet, thus
intensifying both theoreticians' and observers' interest in this peculiar family
of objects.

Fourth, this appeal was still further bolstered by the visual and photographic
detection of a cometary object in the Sun's corona during the solar eclipse on
1882~May~17, just four months before the Great Comet's perihelion passage.  In
retrospect, the event was deemed yet another piece of evidence supporting the
hypothesis of a genetic relationship among sungrazers forged by the process of
recurring fragmentation.

\begin{figure*}
\vspace{0.1cm}
\hspace{-0.2cm}
\centerline{
\scalebox{0.59}{
\includegraphics{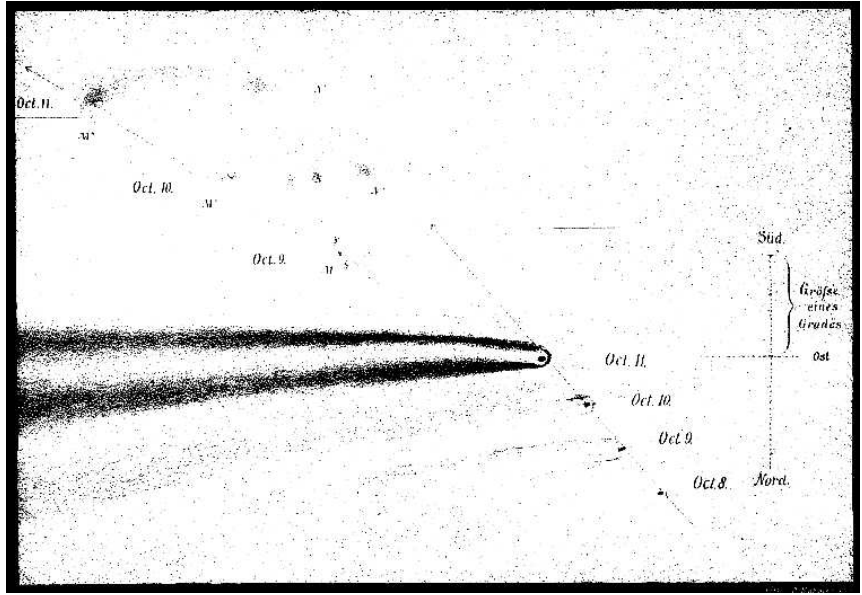}}}
\vspace{0.1cm}
\caption{Schmidt's (1882a) drawing of the Great September Comet and, what he believed,
a single nebulous companion (a.k.a.\ Schmidt's comet).  Plotted are the positions of
the comet's head in the mornings of 1882 October~9--12, the entire tail on October~12,
and a nebulous companion on October~10--12 (Schmidt reckoned the time from Athens local
noon, which explains the disparity in the dates).  Near the right edge are the picture's
orientation (south up, east to the right) and scale (the brace showing a length of
1$^\circ$).{\vspace{0.8cm}}}
\end{figure*}

Fifth, this hypothesis was reinforced to the point of near certainty, when the
nuclear condensation of the Great September Comet itself was observed after
perihelion to consist of up to six separate points of light lined up like a
string of pearls, with the gaps widening with time.  While this compelling piece
of evidence on a near-perihelion breakup of the original nucleus was being
presented literally before the eyes of the observers over a period of months,
an independent but rather elusive corroboration was provided by {\small \bf
nebulous companions}, as explained below.  Unlike the multiple nucleus, they
were harder to observe and, in addition, transient in nature.  Isolated
instances of their sighting, in October 1882, were reported by merely a few
individuals.  They were located at distances of degrees from the head of the
Great September Comet and, with one exception, to the {\it southwest\/} of it.
This type of feature does not seem to have ever been observed with any other
comet and no quantitative explanation of its nature and origin has ever been
offered.  An investigation of this kind is the objective of the present
paper.

\section{Reports and Descriptions of Nebulous Companions} 
In the order of decreasing amount of information provided, the four observers
reporting a feature to the southwest of the main comet's nucleus were Schmidt
(1882a), Hartwig (1883), Barnard (1882, 1883a, 1883b), and Markwick (1883).
Brooks (1883), a fifth observer, remarked on a feature to the east of the
comet.  I now present the major points made by each of them in describing
their findings.{\vspace{-0.2cm}}

\subsection{Schmidt's Account} 
Schmidt (1882a) was the only observer claiming to have seen a nebulous companion
to the southwest of the main comet on more than one occasion.  He reported its
discovery, in the morning of 1882 October~9~UT, as a new comet, noting though that
it traveled alongside the Great September Comet (Kreutz 1882).  The next morning,
October~10.15~UT, Schmidt (1882a) described its appearance in a finder telescope as
a fairly bright, crescent-shaped nebulosity, whose vertex was elongated toward the
west, the uneven wings extending 1$^\circ$ to the south-southwest and 0$^\circ\!$.5
to the east.  The crescent was more condensed near the vertex, fading gradually
toward the horns.  Measured for position was a point in the elongated vertex, which
was more than 3$^\circ$ to the southwest of the main comet's head.

Some 24 hours later, on October 11.13 UT, the companion consisted of two condensations
with wisps and other fainter morphology.  Either condensation was about 7$^\prime$ to
8$^\prime$ in diameter, with no sharp boundary and no nucleus.  The condensations were
some 87$^\prime$ apart, the eastern (seen for the first time) about 20$^\prime$ south
of the western.  The latter, when linked with the elongated vertex on the previous day,
implied a daily motion of $\sim$1$^\circ$ radially away from the main comet's nucleus. 

Lastly, on October 12.13 UT both condensations appeared as very dim and large clouds,
more than 10$^\prime$ across, elongated along a west-east line, with no nucleus
whatsoever, yet noticeably condensed, and connected by a faint bridge of fuzzy
material.  The distance between the condensations was close to 100$^\prime$, the
eastern again a little, about 10$^\prime$, south of the western.  The distance of
the western feature from the main comet's nucleus grew by another degree in 24~hours
in the same direction as before.  Attempts to detect the nebulous companion on the
following mornings failed, in part because of deteriorated observing conditions.

Schmidt presented the results of his observations as offsets from the nucleus of the
main comet as well as in a drawing that is reproduced in Figure~1.  He estimated
that his positional measurements of the nebulous companion were uncertain to
$\pm$0$^\prime\!$.8 on October~10, to $\pm$3$^\prime\!$.1 on October~11, and to
$\pm$2$^\prime\!$.3 on October~12, probably underestimates.  Hind (1882), Oppenheim
(1882), and Zelbr (1882) independently employed the three positions by Schmidt to
compute a parabolic orbit, with mutually incompatible results:\ for example, Hind
determined the argument of perihelion to equal 345$^\circ$ and the perihelion
distance 2.8~{\Rsun}\,, while Oppenheim found for the two elements the values of
108$^\circ$ and 2.0~{\Rsun}\,; Zelbr got, respectively, 80$^\circ$ and 0.84~{\Rsun}
(sic!).  An ephemeris based on Oppenheim's elements placed the object at $\sim$0.8~AU
from the earth on October 10--12, while the main comet was at $\sim$1.4~AU!  Zelbr
noticed that the results were sensitive to the object's assumed geocentric distance
and that an alternative solution offered a prograde, low inclination orbit with a
perihelion distance of about 0.9~AU.  He concluded that unless more data became
available, the object's motion would remain indeterminate.

\subsection{Hartwig's Account} 
Hartwig (1883), as member of a Venus transit expedition, observed a nebulous
companion near the comet from aboard a steamship traveling from Hamburg, Germany, to
Bahia Blanca, Argentina.  At the time of observation, in the morning of October~10,
the ship was cruising in the South Atlantic near a longitude of 44$^\circ$W.  The
nebulous companion appeared as a large nebula, exhibiting a bright nucleus and a
symmetric, widely opened fan-shaped tail.  Attempts to reobserve the feature in
the following mornings, including October~12 under very good conditions, were
unsuccessful.
%

\begin{figure}[t]
\vspace{0.15cm}
\hspace{-0.2cm}
\centerline{
\scalebox{0.98}{
\includegraphics{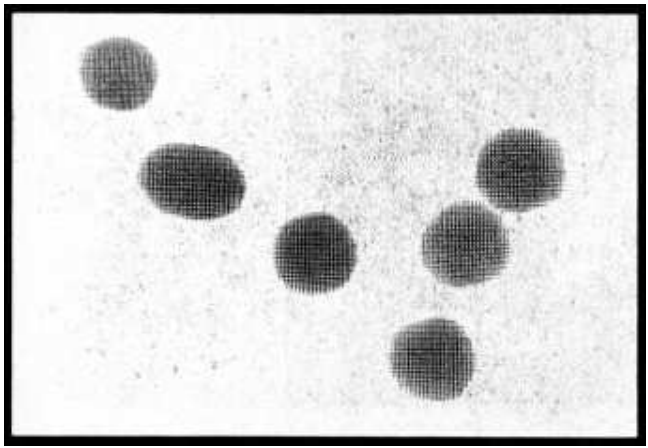}}}
\vspace{0cm}
\caption{Barnard's drawing of six nebulous companions to the Great
September Comet that he detected on October~14.43~UT.  No
scale was presented, but Barnard (1882) indicated that the objects were
about 15$^\prime$ across.  To match approximately his description, south is
to the left and east is up.  The drawing was provided at a request of the
editor of the {\it Sidereal Messenger\/} (Payne 1883).{\vspace{0.4cm}}} 
\end{figure}

\subsection{Barnard's Account} 
Barnard (1882, 1883a) reported that, sweeping to the south of the comet in the
morning of October~14, he first picked up a large mass, 15$^\prime$ in diameter.
It was located between two fainter objects along an east-west line, one in contact,
the other a little farther out on the other side.  Additional objects were found
to the southeast of the trio, including a strongly elongated one.  At least six or
eight of them were within 6$^\circ$ ``south by west'' of the main comet's head.  Each
looked like a telescopic comet slightly condensed to its center, no rapid motion was
apparent.  A drawing, reproduced in Figure~2, was submitted by Barnard in response to
a request by a journal's editor (Payne 1883).  No positional data of higher accuracy
are available.\footnote{It is no secret that Barnard's stellar career had humble
beginnings.  At the time of his 1882 discovery of the nebulous companions (with his
13-cm refractor), he was a 24~year old amateur astronomer with a steep learning curve.
He still lived in his home town of Nashville, Tenn., supporting his wife, mother,
and himself as an employee of a photography studio.  A year earlier, in September
1881, he found his first comet, which boosted his self-esteem and brought not only
recognition by the scientific community but also an award of \$200.  This prize money
was at the time offered by the philanthropist H.\,H.\,Warner for discovery of new
comets by American citizens, once officially recognized by Dr.\,L.\,Swift, Director
of the Rochester Observatory, which was constructed by Warner.  An embarrassing
situation developed when Barnard communicated to Swift his discovery of up to
15~comets (according to Frost 1927; Barnard's own published account quoted above
referred to a smaller number) near the Great September Comet, after he had spent
a restless night filled with dreams of countless comets in the field of view of
his telescope.  Swift may have deemed it unconscionable to ask Warner for
thousands of dollars to reward Barnard's dream-triggered discovery bonanza, thereby
never distributing the news of the achievement.  It was up to Barnard himself to
publish his findings, which he did without mentioning the dreams.}  A hazy sky
prevented Barnard from searching for the objects the following morning, and he
never saw them again; he was very puzzled by their transient nature (Barnard 1883b).

\subsection{Markwick's Account} 
Markwick (1883) appended his report on the Great Comet from the morning of October~5
by briefly noting that south preceding (i.e., to the southwest of) the comet's head
and about 1$^\circ\!$.5 distant were two wisps of nebulous light.  He admitted that
he could not say whether they had anything to do with the main comet, and he was
unable to recover them in the following mornings.
 
\subsection{Brooks' Account} 
While one could speculate that the nebulous companions in the southwest, described
in Sections~2.1 to 2.4, may somehow have been related to one another, the cometary
mass detected by Brooks' (1883) 8$^\circ$ to the east of the main comet's head in
the morning of October 21 was unquestionably a feature of different nature.  Brooks
unfortunately offered only a limited amount of information; he reported that the
mass was nearly 2$^\circ$ long and had a slight condensation near its larger end,
which pointed toward the Sun.  Brooks confirmed his discovery by a second (and last)
observation the following morning, when the object was fainter, smaller by at least
0$^\circ\!$.5, and the condensation was barely perceptible.  The amount of available
information is insufficient to warrant a study of this feature.

\subsection{Summary of Companions' Positional Data} 
Table 1 presents the positional data on the nebulous companions that I collected
from reports by the five observers.  For each time of observation, converted to UT,
I list the polar coordinates of the nebulous companions' offsets from the head of
the Great Comet.  With the exception of Schmidt (1882a) and Hartwig (1883), the
position-angle data were reported by the observer, albeit in approximate terms
only.  Schmidt listed both the separation distance and the offsets in declination
and right ascension for the western part of the observed feature.  The tabulated
position angle was derived by the author from the offsets in right ascension and
declination converted to the equinox of J2000.  The offsets of the eastern part
were handled similarly, after the differences in right ascension and declination
from the western part were first added to the latter's offsets from the head of the
main comet.  Hartwig's (1883) published equatorial coordinates for the ``middle''
of the nebulosity were precessed to J2000 and employed to compute the offsets from
the main comet.

\begin{table}[t]
\vspace{-4.22cm}
\hspace{5.25cm}
\centerline{
\scalebox{1.01}{
\includegraphics{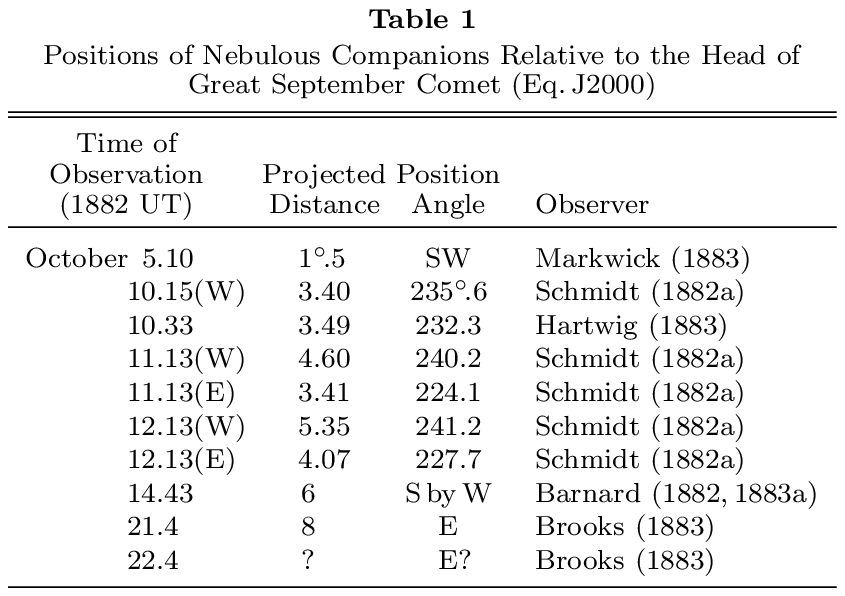}}}
\vspace{-19.1cm}
\end{table}

\section{Nebulous Companions as the Comet's Fragmentation Products} 
To start with, I subscribe to the prevailing view that the nebulous companions
were genetically related to the Great September Comet, or, put another way, they
were its fragments.  Of the observers only Markwick (1883) appears to have been
noncommittal on this issue.  In the following I focus on the nebulous companions
situated to the southwest of the main comet.  The point one should feel a little
uneasy about --- and whose explanation should follow from the solution --- is
that when it comes to the detection of a nebulous companion on a particular day,
there are apparent contradictions among the observers.  For example, even though
both Hartwig and Schmidt reported a nebulous companion at similar positions on
October~10~UT, Hartwig, unlike Schmidt, could not recover it under very favorable
conditions on the 12th.  On the other hand, Schmidt failed to see any trace of a
nebulous object on seven consecutive mornings, October~13--19, even though the
observing conditions were very favorable on four of them and Barnard detected a
swarm of such objects on the 14th.  Given the spread of the observers' sites in
the geographic longitude, it almost looks as though the features were only visible
over periods as short as a fraction of the day, bringing Barnard's concern on
their transient nature to an extreme.  Also peculiar is the circumstance that ---
contrary to Schmidt's overtly stated expectations --- no reports whatsoever were
coming from prestigious southern observatories (such as Cape or C\'ordoba), more
favorably located than Schmidt (Athens, Greece), Barnard (Nashville, Tenn.), or
Brooks (Phelps, N.Y.).  

In order to get an insight into the process that landed these objects degrees to the
southwest of the main comet, I focus on the western condensation of the nebulous
companion measured by Schmidt (1882a) on October~10 and 12.  If his measurements
should indeed refer to the same object, the derived motion should offer constraints
on models.  Predicating this investigation on the research results for split comets
(e.g., Sekanina 1982, Boehnhardt 2004), I accept that the separation distance between
a fragment and its parent at a given time is governed by (i)~the fragmentation time;
(ii)~the acquired separation velocity, typically on the order of 1~m~s$^{-1}$, which
is independent or nearly independent of heliocentric distance; and (iii)~the
differential nongravitational acceleration that the fragment is subjected to after
separation.  This acceleration is believed to be outgassing-driven (mainly due to
the sublimation of water ice), thereby varying approximately as an inverse square
of heliocentric distance when less than about 1~AU from the Sun.  Its magnitude for
fragments of known split comets is usually in a range of 10$^{-5}$ to 10$^{-4}$ the
Sun's gravitational acceleration (which equals 0.593~cm~s$^{-2}$ or \mbox{$29.6 \times
\! 10^{-5}$\,AU day$^{-2}$} at 1~AU from the Sun), depending on the fragment's mass,
activity, etc.  The acceleration correlates with the fragment's lifespan, being higher
for short-lived objects and lower for persistent ones.  As a rule, the acceleration
is directed essentially away from the Sun and its effects dominate those of the
separation velocity.  A major exception is presented by the population of SOHO dwarf
Kreutz sungrazers, whose examination by Sekanina \& Kracht (2015) demonstrated that
(i)~the {\it normal\/} component of the nongravitational acceleration was comparable
in magnitude to the radial component and (ii)~the acceleration's overall magnitude
was in the range of 0.02 to nearly 1~Sun's gravitational acceleration (sic!) between
8 and 15~{\Rsun} from the Sun.

An important dynamical property of fragments of split comets whose motions are
dominated by the {\it radial\/} nongravitational acceleration is that, relative
to the primary (most massive) fragment, the motions of secondary (less massive)
fragments are directionally constrained:\ their positions in the orbital plane
are restricted to a sector of less than 180$^\circ$, subtended by the prolonged
radius vector, {\small \boldmath $RV$} (the anti-solar direction) and the vector
of the negative orbital-velocity vector, {\small \boldmath $-V$} (the direction
behind the comet).  To verify compliance with this constraint is trivial.

To subject the Schmidt nebulous companion to this test, I list the position angles,
measured from the nucleus of the Great September Comet, in Table~2:\ for the radius
vector in column~4, for the negative velocity vector in column~5.  Because the orbit's
curvature at the relevant heliocentric distances is very small, the two vectors make
an angle of nearly 180$^\circ$, yet it is the northern sector that is slightly less
than 180$^\circ$ wide.  For example, for October~10.15~UT a nebulous companion would
satisfy the condition if its position angle were smaller than 84$^\circ\!$.8 or
greater than 267$^\circ$.  Since the Schmidt nebulous companion was near position
angle of 240$^\circ$, it did not satisfy the condition and its motion was obviously
not dominated by the radial nongravitational acceleration, regardless of the
fragmentation time.

\begin{table}[t]
\vspace{-4.2cm}
\hspace{5.25cm}
\centerline{
\scalebox{1}{
\includegraphics{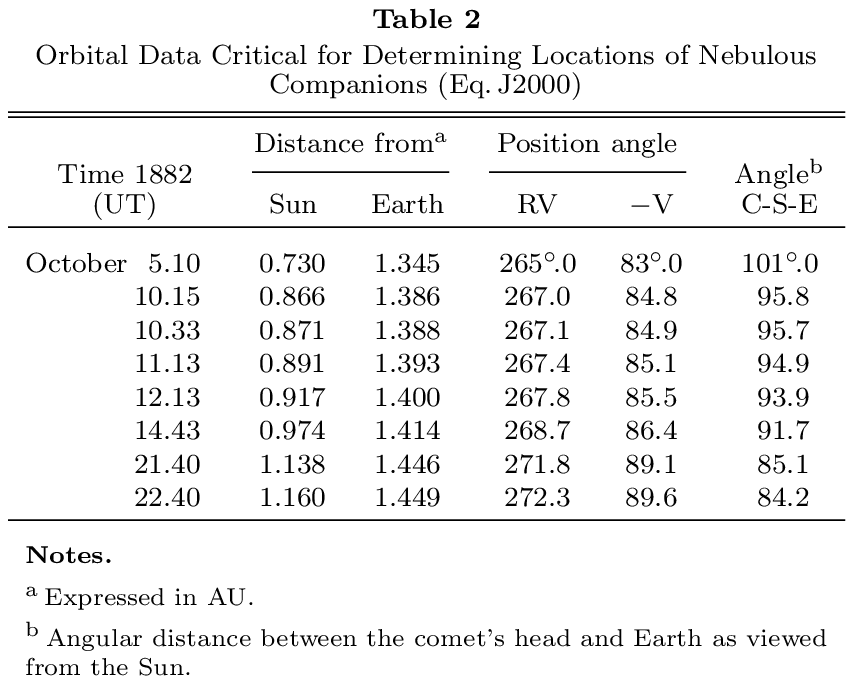}}}
\vspace{-18.3cm}
\end{table}

The obvious alternative is to consider the nebulous companion's motion determined
mainly by a separation velocity.  Schmidt's observations of the nebulous companion's
western condensation suggest a motion rate of about 1$^\circ$ per day at a geocentric
distance of 1.4~AU, assuming the nebulous companion was in the general proximity
of the main comet.  This transforms to an average {\it projected \/} velocity
relative to the main comet of approximately 40~km~s$^{-1}$.  Even before accounting
for a foreshortening effect, this is more than four orders of magnitude higher than
an expected relative velocity, ruling out the separation velocity as the trigger.

\section{Nongravitational Acceleration Normal to the Orbit Plane} 

Another option is to contemplate that the nebulous companion's motion relative to the
main comet was governed by a nongravitational acceleration with a strong contribution
from the component normal to the orbital plane, a scenario emulating the behavior of
the SOHO Kreutz sungrazers.  To examine this case, I first consider a fragment that
is moving with the comet except for its increasing distance from the orbital plane.
I consider a right-handed orthogonal rotating heliocentric coordinate system in which
the $z$-axis points to the comet's orbital pole, the $x$-axis to the comet's nucleus,
and the $y$-axis completes the system.  The position of the main comet's nucleus at
time $t$ is determined by the coordinates \mbox{$\{r(t), 0, 0\}$}, while the nebulous
companion, subjected to a nongravitational acceleration normal to the orbital plane, is
assumed to be located at a point \mbox{\{$r(t), 0, z(t)$\}}.  The task is to determine
the distance $z(t)$ from the orbital plane as a function of time.  To make the problem
easily tractable, I use some (reasonable) approximations.

Let a nebulous companion be released from the comet's nucleus after perihelion, at
time $t_{\rm frg}$, when its heliocentric distance is \mbox{$r_{\rm frg} = r(t_{\rm
frg})$}.  Let the orbit of the main comet be approximated by a parabola, so that
the relation between time $t$ (in days) and heliocentric distance $r(t)$ (in AU) is
expressed by
\begin{equation}
t = t_\pi + {\displaystyle \frac{\sqrt{2}}{3k}} (r\!+\!2q) \sqrt{r\!-\!q}, 
\end{equation}
where $t_\pi$ is the time (in days) of the main comet's perihelion passage
and $k$ is the Gaussian{\vspace{-0.05cm}} gravitational constant,
\mbox{$k = 0.0172021$ AU$^{\frac{3}{2}}$ day$^{-1}$}.  Let the nebulous
companion be subjected to a nongravitational acceleration normal to the
orbital plane, whose variation with heliocentric distance follows an inverse
square power law.  As a result, the nebulous companion begins to recede from the
comet in a direction perpendicular to the orbital plane with a systematically
increasing velocity $\dot{z}(t)$, 
\begin{equation}
\dot{z}(t) = \int_{t_{\rm frg}}^{t} \!\!\! \frac{k^2\Gamma_{\!\rm N}}{r^2}\,dt, 
\end{equation}
where $k^2/r^2$ is the Sun's gravitational acceleration at $r$ and $\Gamma_{\rm N}$ is
a dimensionless constant that determines the normal component of the nongravitational
acceleration as a fraction of the Sun's gravitational acceleration.  It is noted that
\mbox{$\dot{z}(t_{\rm frg}) = 0$} and that the{\vspace{-0.05cm}} denominator in Equation~(2)
should strictly read \mbox{$r^2 \!+\! z^2$}; the approximation is acceptable as long as
\mbox{$z \ll r$}.  The distance from the comet (and the orbital plane) at time $t$ is
\begin{equation}
z(t) = \!\int_{t_{\rm frg}}^{t} \!\!\!\! \dot{z}(t) \, dt. 
\end{equation}
One can now proceed in one of two possible ways.  One is to use Kepler's second law
and substitute true anomaly for time as the integration variable.  
Followed below is the other option, in which $dt$ is expressed in
terms of $dr$ by differentiating Equation~(1),
\begin{equation}
dt = \frac{r}{k\sqrt{2}\sqrt{r-q}} \, dr.                                       
\end{equation}
The integration of the velocity expression (2) gives for $z(t)$ from Equation~(3)
with help of the relation (4):
\begin{eqnarray}
z(t) & = & \frac{\Gamma_{\!\rm N}}{\sqrt{q}} \! \int_{r_{\rm frg}}^{r} \!\!
 \frac{r}{\sqrt{r-q}} \! \left[ \arcsin \!\sqrt{q/r_{\rm frg}} - \!\arcsin \!\sqrt{q/r}
 \right] \! dr \nonumber \\
     & = & 2 q \Gamma_{\!\rm N} \!\!\int_{\sqrt{q/r}}^{\sqrt{q/r_{\rm frg}}}
 \frac{\arcsin \sqrt{q/r_{\rm frg}} - \arcsin u}{u^4 \sqrt{1 \!-\! u^2}} \, du. 
\end{eqnarray}
Since the integration of the expression in closed form is rather convoluted, I prefer
to expand the functions into polynomials,
\begin{eqnarray}
\arcsin u & = & u + {\textstyle \frac{1}{6}} u^3 + \ldots, \nonumber \\
(1 \!-\! u^2)^{-\frac{1}{2}} & = & 1 + {\textstyle \frac{1}{2}} u^2 + {\textstyle
 \frac{3}{8}} u^4 + \ldots,                                               
\end{eqnarray}
and, assuming that \mbox{$q \ll r_{\rm frg} < r$}, neglect{\vspace{-0.07cm}} all terms higher
than $u^2$.  Substituting \mbox{$\Re = \sqrt{r/r_{\rm frg}} > 1$}, I obtain a solution
\begin{equation}
z(t) = {\textstyle \frac{1}{3}} \Gamma_{\!\rm N} \left[ r_{\rm frg} F(\Re) + 3q G(\Re)
 \right],
\end{equation}
where
\begin{eqnarray}
F(\Re) & = & (\Re - 1)^2 (2\Re + 1), \nonumber \\
G(\Re) & = & (\Re - 1) - \ln \Re.                                               
\end{eqnarray}
The ratio $G(\Re)/F(\Re)$ decreases from $\frac{1}{6}$ when \mbox{$\Re \!\rightarrow\!
1$} to zero when \mbox{$\Re \!\rightarrow\! \infty$}, so that the second term in
Equation~(7) always contributes only a fraction of the first.  To test the range of
$\Gamma_{\!\rm N}$ required to fit the data in Table~1, I first consider a case of an
angular distance of 4$^\circ$ at a heliocentric distance of 0.9~AU and a geocentric
distance of 1.4~AU; the required nongravitational accelerations $\Gamma_{\!\rm N}$
are found to be high, from 10~percent of the solar gravitational acceleration for
a fragment separating at a heliocentric distance 0.1~AU (1~day after perihelion,
September 18.7~UT) to 7.3~times the solar gravitational acceleration for a fragment
separating at 0.7~AU from the Sun (16.3~days after perihelion, October 4.0~UT)!

This exercise offers only a crude estimate, because the projection conditions have
not as yet been considered.  In addition, a nebulous companion's motion in a strictly
perpendicular direction to the orbital plane is unrealistic, as documented by a
simple calculation of the direction of the projected normal to the orbital
plane.  For the times of Schmidt's observations of the nebulous companion, the
position angle of the northern orbital pole was essentially constant, equaling
166$^\circ$.  The nebulous companion surely did not travel along the normal to the
orbital plane, given the difference of about 70$^\circ$ between the observed and
expected position angles.

\begin{figure*}
\vspace{-6.7cm}
\hspace{-0.35cm}
\centerline{
\scalebox{0.81}{
\includegraphics{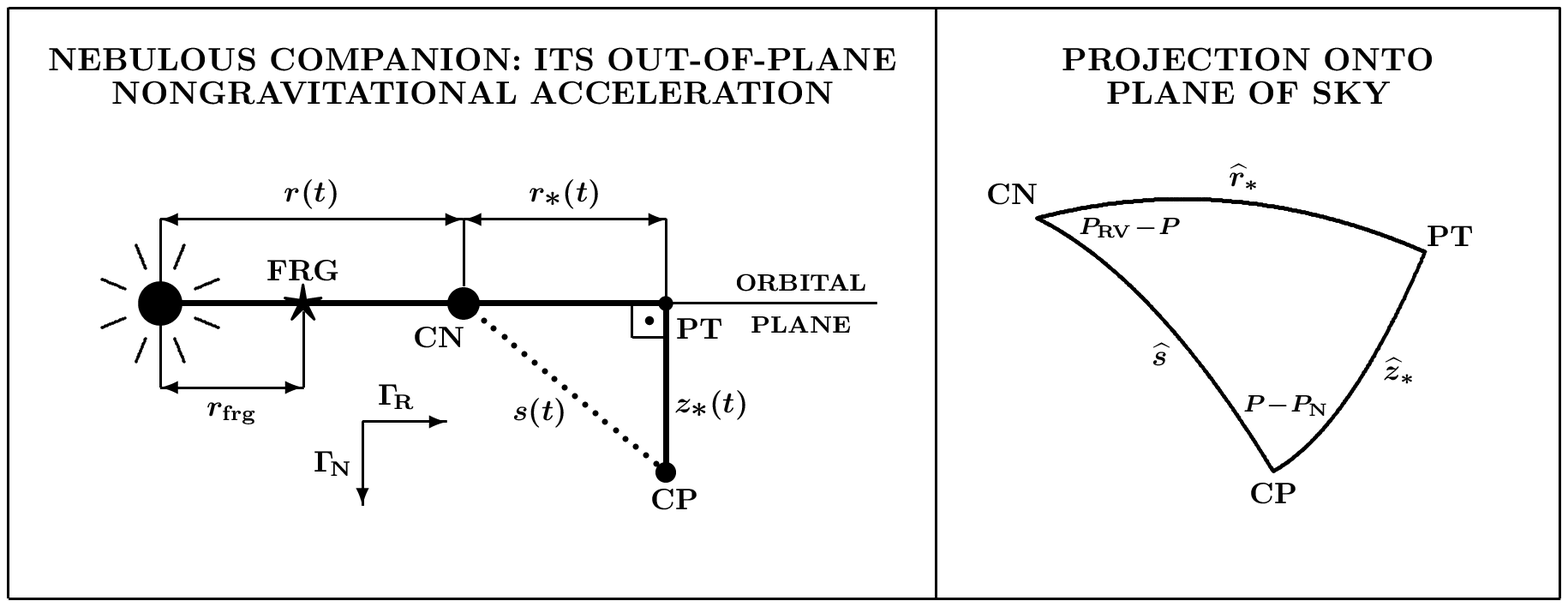}}}
\vspace{-11.4cm}
\caption{Nebulous companion to comet C/1882 R1 subjected to an outgassing-driven
nongravitational acceleration, whose radial component, measured in units of
the Sun's gravitational acceleration, is $\Gamma_{\!\rm R}$ and the normal
component is $\Gamma_{\!\rm N}$.  {\sf Left}:\ View from a point in the orbital
plane. The Sun is to the left, {\scriptsize \bf FRG} is the point of separation,
from the main comet, of the fragment whose debris the nebulous companion
represents. {\scriptsize \bf CN} and {\scriptsize \bf CP} are the positions of,
respectively, the main comet's head and the nebulous companion at the time of
observation, $t$.  {\scriptsize \bf PT} is the companion's footprint in the orbital
plane.  The{\vspace{-0.035cm}} heliocentric distance of the point of separation is
$r_{\rm frg}$, while $r(t)$ is the comet's heliocentric distance at $t$.  The distance
of the nebulous companion from the main comet at time $t$ is $s(t)$,
from the orbital plane $z_\ast(t)$. The distance of the companion's footprint from
the main comet at time $t$ is $r_\ast(t)$. {\sf Right}:\ In projection onto the
plane of the sky.  {\scriptsize \bf CN}, {\scriptsize \bf CP}, and {\scriptsize
\bf PT} refer to the same points as on the left. The angular distance of the
companion from the main comet is $\widehat{s}$, from its footprint in the orbital
plane is $\widehat{z}_\ast$, and the angular distance of the footprint point from
the main comet is $\widehat{r}_\ast$.  The position angle of the companion from
the main comet is $P$, the position angle of the radius vector measured at the
main comet is $P_{\rm RV}$, and the position angle of the normal to the orbital
plane measured at the companion is $P_{\rm N}$.  The measured quantities are
$P$ and $\widehat{s}$, while $P_{\rm RV}$ and $P_{\rm N}$ can be derived from
the orbital elements and the companion's position.{\vspace{0.8cm}}}
\end{figure*}

\section{Two-Dimensional Model for a Nebulous Companion's Motion} 
Even though the nongravitational acceleration in a direction normal to the orbital
plane was positionally the most consequential effect, the in-depth treatment
of the motions of the eight SOHO sungrazers by Sekanina \& Kracht (2015) showed
that the normal component was generally on a par with the acceleration's radial
component.  Consistent with this finding, I revise the scenario from the foregoing
section, requiring now that the nebulous companion's ``footprint'' in the orbital
plane be not the comet's nucleus but a hypothetical particle that separated from the
nucleus at time $t_{\rm frg}$ under a {\it radial\/} nongravitational acceleration
of $\Gamma_{\!\rm R}$.  Because of the nature of sungrazing orbits, this particle was
moving essentially along the radius vector (except very near the Sun, a case not
contemplated here), receding from the nucleus to a distance of $r_\ast(t)$ over a
period of time \mbox{$t \!-\! t_{\rm frg}$}.  Accordingly, in the system introduced
in Section~4, the coordinates of the nebulous companion's footprint in the orbital
plane at time $t$ are \mbox{$\{r \!+\! r_\ast(t), 0, 0\}$} and the coordinates of
the nebulous companion itself are $\{r \!+\! r_\ast(t), 0, z_\ast(t)\}$, where,
in analogy to Equation~(7),
\begin{equation}
r_\ast(t) = {\textstyle \frac{1}{3}} \Gamma_{\!\rm R} [r_{\rm frg} F(\Re)
  + 3q G(\Re) ]                                                             
\end{equation}
and the $z_\ast$ coordinate of the nebulous companion is then
\begin{equation}
z_\ast(t) = {\textstyle \frac{1}{3}} \Gamma_{\!\rm N} [r_{\rm frg} F(\Re_\ast) + 3q
 G(\Re_\ast)],                                                              
\end{equation}
where \mbox{$\Re_\ast = \sqrt{(r \!+\! r_\ast)/r_{\rm frg}}$}.
%

The task now is to find out under what circumstances could this simple model fit the
motion of the nebulous companion observed by Schmidt between October~10 and 12~UT.
I ignore the middle observation, on October~11~UT, which was described by Schmidt as
the most uncertain of the three anyway.  The line of attack is described schematically
in Figure~3.  The left panel shows a view of the post-perihelion branch of the orbit
from a point in the orbital plane.  From the left to the right are the Sun; the point
of separation of the nebulous companion from the comet's nucleus (FRG); the comet's
position (CN) at time $t$; and the nebulous companion's position (CP) and footprint in
the orbital plane (PT) at time $t$.  The distances $r_{\rm frg}$, $r(t)$, $r_\ast(t)$,
$z_\ast(t)$, and the separation distance of CP from CN, $s(t)$, are marked.  The right
panel transfers the picture from space onto the plane of the sky.  The three vertices
of the spherical triangle, CN, PT, and CP, are pairwise connected by great circular
arcs:\ $\widehat{s}$ links CN with CP, $\widehat{r}_\ast$ links CN with PT, and
$\widehat{z}_\ast$ links CP with PT.  One knows two of the three angles:\ (i)~the angle
at the vertex CN equals the difference between the position angle of the prolonged
radius vector, $P_{\rm RV}$, and the position angle of the nebulous companion measured
from the comet's nucleus, $P$; and (ii)~the angle at the vertex CP equals the
difference between $P$ and the position angle of the normal to the orbital plane,
$P_{\rm N}$.  The arcs $\widehat{r}_\ast$ and $\widehat{z}_\ast$ are then determined
from
\begin{eqnarray}
\sin \widehat{r}_\ast & = & \kappa \sin (P \!-\! P_{\rm N}) \nonumber \\
\sin \widehat{z}_\ast & = & \kappa \sin (P_{\rm RV} \!-\! P),              
\end{eqnarray}
where
\begin{eqnarray}
\kappa = \sin \widehat{s} \left\{ \! \raisebox{0ex}[1ex][1ex]{1} - \left[ \cos (P_{\rm RV}
 \!-\! P) \cos (P \!-\! P_{\rm N}) \right. \right. \nonumber \\[-0.2cm]
 \left.\left.  - \cos \widehat{s} \, \sin (P_{\rm RV} \!-\! P) \sin (P \!-\! P_{\rm N\!})
 \right]^2 \right\}^{\!-\frac{1}{2}} \!\!.                                             
\end{eqnarray}
Next, the angular distances $\widehat{r}_\ast$ and $\widehat{z}_\ast$, which I
express in degrees, need to be converted to the linear distances $r_\ast$ and
$z_\ast$ that will be reckoned in million km.  This requires to account for the
effects of geocentric distance and projection foreshortening.  Given that 1$^\circ$
projected onto the plane of the sky equals 2.611~million km at 1~AU from the earth,
the conversion is accomplished by the relations:
\begin{eqnarray}
r_\ast(t) & = & 2.611 \, \Delta(t) \,f_{\rm RV}\!(t) \:\widehat{r}_\ast, \nonumber \\[0.1cm]
z_\ast(t) & = & 2.611 \,\Delta(t) \,f_{\rm N}(t) \:\widehat{z}_\ast,              
\end{eqnarray}
where $\Delta(t)$ is the comet's geocentric distance at time $t$ listed in Table~2
(the distance between the nucleus and the nebulous companion being neglected) and
$f_{\rm RV}(t)$ and $f_{\rm N}(t)$ are the projection foreshortening corrections
along the radius vector and the normal to the orbital plane, respectively.
Numerically the factor $f_{\rm RV}$ is related to the angle $\psi_{\rm RV}(t)$
between the line of sight and the plane normal to the radius vector by
\begin{equation}
f_{\rm RV}(t) = \sec \psi_{\rm RV}(t) \geq 1.                               
\end{equation}
This angle is equal to \mbox{$\psi_{\rm RV} = |90^\circ \!-\! ($C-S-E)$|$}, where
\mbox{(C-S-E)} is the angle Comet-Sun-Earth listed in Table~2.  At Schmidt's October
10--12 observations $\psi_{\rm RV}$ varied between 3$^\circ\!$.9 and 5$^\circ\!$.8,
so that $f_{\rm RV}$ did not exceed 1.005.  The similarly defined angle $\psi_{\rm N}$
varied from 10$^\circ\!$.5 on October~10 to 11$^\circ\!$.2 on October~12, implying
$f_{\rm N}$ in a range from 1.017 to 1.019.

The objective is to find a solution providing the time of the nebulous companion's
separation from the comet's nucleus that would simultaneously fit Schmidt's
observations from October~10.15~UT, referred to below as time $t_1$, and October
12.13~UT, or time $t_2$.  More specifically, the values of $r_\ast(t_1)$,
$z_\ast(t_1)$, $r_\ast(t_2)$, and $z_\ast(t_2)$ derived from the observed data
via Equations~(13) should equal the values computed from Equations~(9) and (10),
respectively.  Since the latter depend on the nongravitational acceleration's
components $\Gamma_{\!\rm R}$ and $\Gamma_{\!\rm N}$, which are, next to $r_{\rm
frg}$, additional free parameters, it is only necessary that the ratios
$r_\ast(t_2)/r_\ast(t_1)$ and, simultaneously, $z_\ast(t_2)/z_\ast(t_1)$ from the
proposed scenario match those derived directly from the observations.

Solving this problem is convolved not only because the search has to be conducted
by trial and error, but primarily because the observations are burdened by large
errors.  The uncertainty is of two kinds:\ one is triggered by a great difficulty to
bisect the brightest point of a fuzzy feature of large dimensions, a problem that
Schmidt kept complaining about; the other, even more severe, is prompted by the
nebulous companion's ever changing morphology, which makes the position of the
brightest spot refer to different parts of the feature over time; the observer's
measurements of the nebulous companion are then rather misleading.  The bottom
line is that the search for a solution should aim at examining a fit to not only
the reported position, but to a class of positions in its general proximity to
determine an optimum case.

Armed with this strategy, I found that the rate of increasing separation of the
nebulous companion from the main comet is fitted best by assuming that the nebulous
companion broke off at a heliocentric distance of 0.665~AU, one week before Schmidt's
October~10 observation.  The nebulous companion's observed motion in the orbit's
plane was matched very well, but moderate corrections were required in the direction
normal to the plane, 12$^\prime$ to the north on the 10th and 26$^\prime$ to the
south on the 12th, equivalent to changes of a few degrees in the position angle.
The results in Table~3 show that fitting the motion of the nebulous companion
relative to the comet entailed an extremely high nongravitational acceleration in
the radial direction, more than five times higher than the Sun's gravitational
acceleration.  In the normal direction the effect was smaller, but still exceeding
the Sun's acceleration.

The source of these huge forces is unknown and suspicious, and their existence dubious.
It is easy to show that they cannot come from the sublimation of ices by the nebulous
companion's material.  An order-of-magnitude estimate is obtained from the well-known
conservation-of-momentum condition:\ If ${\cal M}$ is the{\vspace{-0.07cm}} nebulous
companion's mass at time $t$, $\dot{\cal M}_{\rm subl}$ an average mass-loss rate by
outgassing during the nebulous companion's lifetime, between $t$ and \mbox{$t \!+\!
\tau_{\rm life}$}, $V_{\rm subl}$ a collimated sublimation velocity (not exceeding
0.36~km~s$^{-1}$), and $\gamma_{\rm subl}$ an average outgassing-driven nongravitational
acceleration,{\vspace{-0.07cm}} the momentum condition, \mbox{${\cal M} \gamma_{\rm
subl} = V_{\rm subl} \dot{\cal M}_{\rm subl}$}, implies for the nebulous companion's
sublimation lifetime
\begin{equation}
\tau_{\rm life} = \frac{\cal M}{\dot{\cal M}_{\rm subl}} = \frac{V_{\rm
  subl}}{\gamma_{\rm subl}}.                                                 
\end{equation}
With the acceleration data from Table 3 near 0.9~AU, I find \mbox{$\gamma_{\rm subl}
= 4.15$ cm s$^{-2}$}, and the expected sublimation lifetime is
\begin{equation}
\tau_{\rm life} < 0.1 \; {\rm day},                                        
\end{equation}
which is two orders of magnitude shorter than the companion's time of flight from
Table~3.  The obvious difficulty with this scenario is to explain the nebulous
companion's enormous relative velocity implied by its motion of about 1$^\circ$
per day that Schmidt (1882a) specifically noted.  This angular rate is equivalent
to a projected linear velocity of about 42~km~s$^{-1}$ if the nebulous companion is
crudely at the same geocentric distance as the main comet.  This enormous relative
velocity ought to be reached over a very short period of time, as the nebulous
companion would have otherwise been brighter earlier and should have been detected
sooner by many observers on numerous occasions.

From the standpoint of the proposed hypothesis, the distances $r_\ast$ and $z_\ast$
are tightly constrained by their ratio from the two dates.  For example, the ratio
$r_\ast$(Oct\,12)/$r_\ast$(Oct\,10) ought to be 1.55; it equals only 1.12 and 1.28
when the separation time $r_{\rm frg}$ is 0.1~AU and 0.5~AU, respectively.  It
equals 1.55 {\it only\/} when \mbox{$r_{\rm frg} = 0.665$ AU}.  Similarly the ratio
of 1.76 for $z_\ast$.  There is no escape from the constraint of recent separation
and thus of unrealistically high nongravitational accelerations.

\begin{table*}
\vspace{-4.2cm}
\hspace{0.5cm}
\centerline{
\scalebox{1}{
\includegraphics{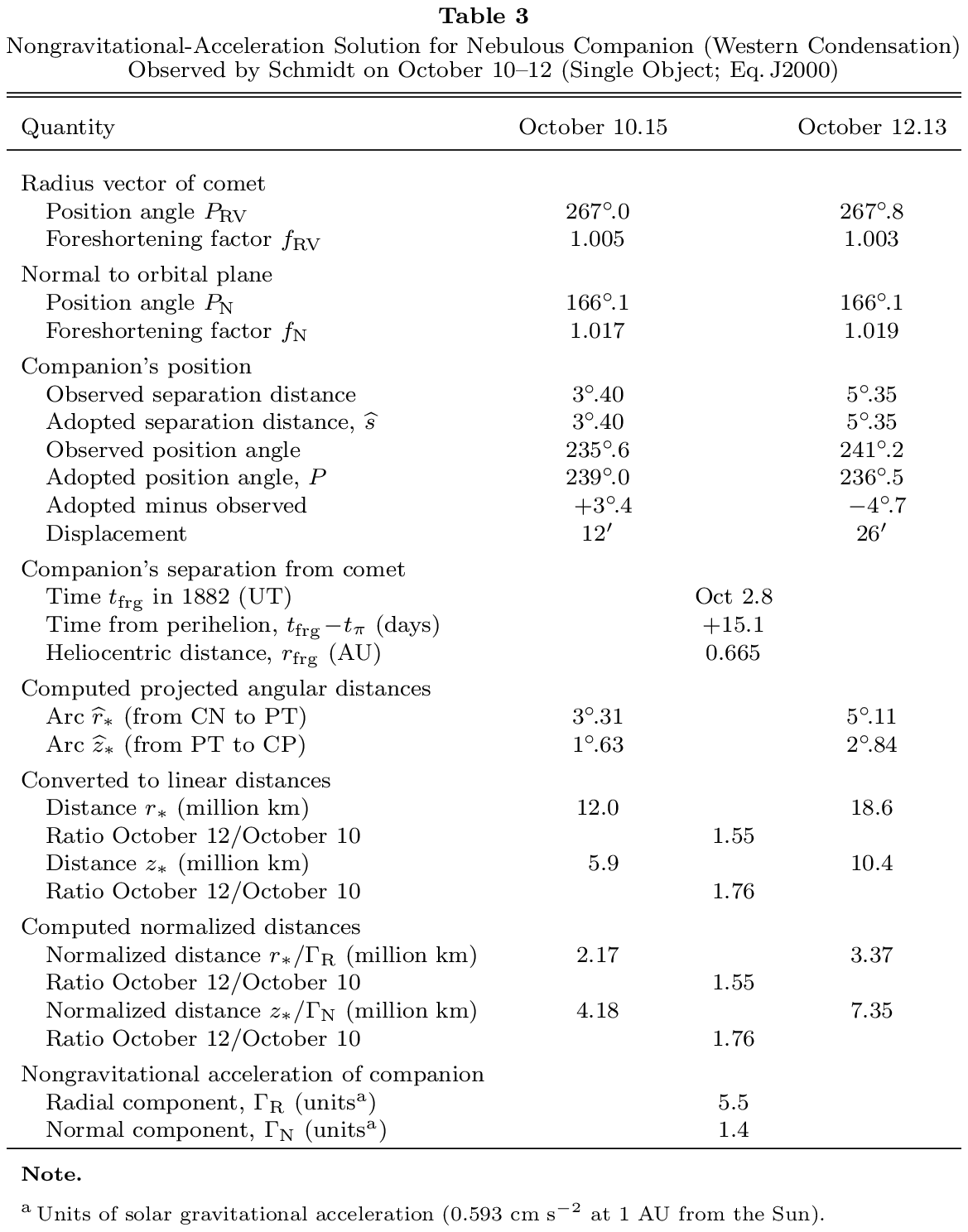}}}
\vspace{-10.1cm}
\end{table*}

\section{Introducing the Separation Velocity} 
In an effort to mitigate these disparities, I now assume that the nebulous companion
departed the comet with a separation velocity, whose radial and normal components
were $V_{\rm R}$ and $V_{\rm N}$, respectively, before it became subject to the
nongravitational acceleration.  This premise requires that the expressions for
$r_\ast$ and $z_\ast$ be corrected to include the respective velocity contributions
as functions of heliocentric distance.

The expressions for the radial and normal components are again of the same type;
I call $r_0$ the integrated effect of the radial component of the separate velocity,
$V_{\rm R}$; and $z_0$, the integrated effect of the normal component, $V_{\rm N}$,
\begin{eqnarray}
r_0(t) & = & \int_{t_{\rm frg}}^{t} \!\!\!\! V_{\rm R} \, dt
         = \frac{V_{\rm R}}{k \sqrt{2}} \int_{r_{\rm frg}}^{r}
           \!\frac{r}{\sqrt{r \!-\! q}} \, dr, \nonumber \\[0.1cm]
z_0(t) & = & \int_{t_{\rm frg}}^{t} \!\!\!\! V_{\rm N} \, dt
         = \frac{V_{\rm N}}{k \sqrt{2}} \int_{r_{\rm frg}}^{r}
           \!\frac{r}{\sqrt{r \!-\! q}} \, dr,            
\end{eqnarray}
where once again I use the parabolic approximation.  Integrating the expression
for $r_0(t)$ I find
\begin{eqnarray}
r_0(t) & = & \frac{V_{\rm R}\sqrt{2}}{3k} \!\left[ \raisebox{0ex}[3ex][3ex]{}
 r^{\frac{3}{2}} \!\left(1\!-\!{\displaystyle \frac{q}{r}} \right)^{\!\frac{1}{2}}
 \!\!\left(1\!+\!{\displaystyle \frac{2q}{r}}\right) \right.
 \nonumber \\
 & & \left. - r_{\rm frg}^{\frac{3}{2}} \!\left( \!1\!-\!{\displaystyle
 \frac{q}{r_{\rm frg}}} \!\right)^{\!\frac{1}{2}} \!\!\left( \!1\!+\!{\displaystyle
 \frac{2q}{r_{\rm frg}}} \!\right) \!\right].                                                                  
\end{eqnarray}

\begin{table*}
\vspace{-4.2cm}
\hspace{0.5cm}
\centerline{
\scalebox{1}{
\includegraphics{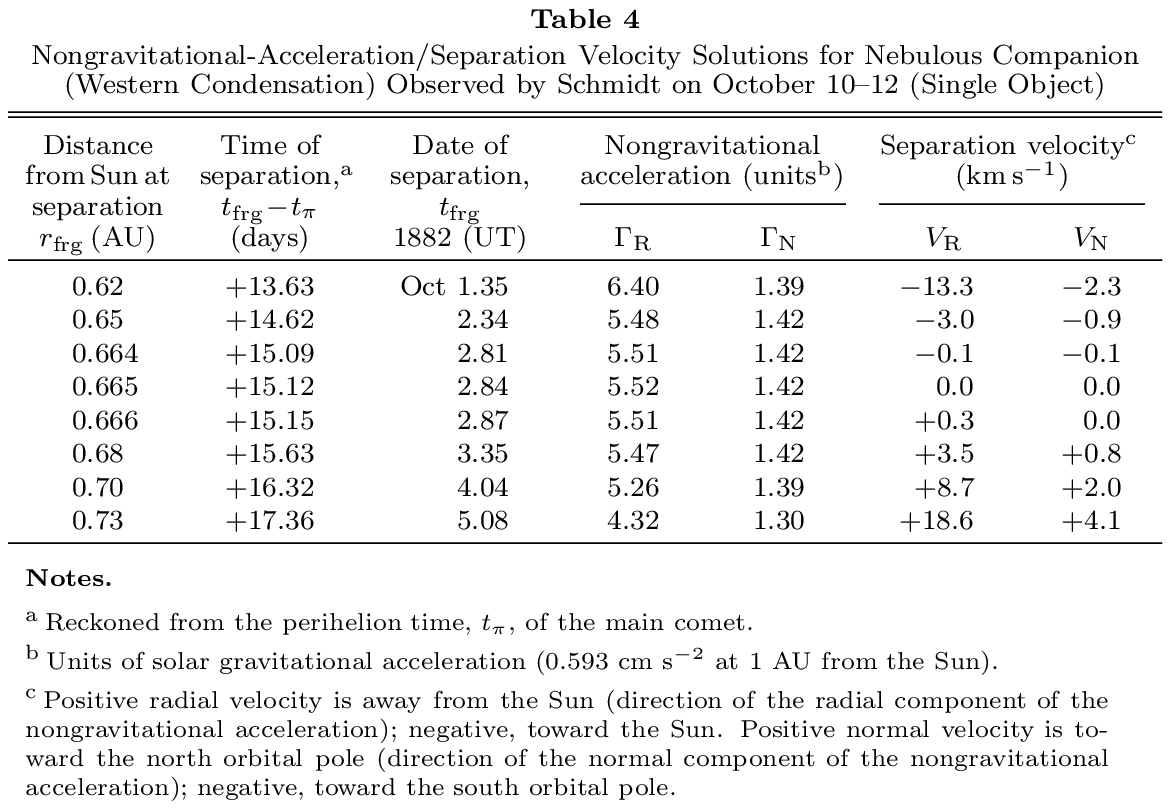}}}
\vspace{-16.8cm}
\end{table*}

\noindent Writing the square root expressions as series and neglecting the
$(q/r_{\rm frg})$ terms of powers 2 and higher, one has:
\begin{equation}
r_0(t) = \frac{V_{\rm R}\sqrt{2r_{\rm frg}}}{3k} \!\left[ r_{\rm frg} (\Re^3 \!-\! 1)
 + {\textstyle \frac{3}{2}} q (\Re \!-\! 1) \right],                        
\end{equation}
where \mbox{$\Re = \sqrt{r(t)/r_{\rm frg}}$}. The heliocentric distance of the
nebulous companion's footprint in the orbital plane at time $t$ is now equal to
\mbox{$r(t) \!+\! r_\ast(t) \!+\! r_0(t)$}.  Replacing in (19) $V_{\rm R}$ with
$V_{\rm N}$ and{\vspace{-0.06cm}} the ratio $\Re(t)$ with \mbox{$\Re_\ast(t) =
\sqrt{[r(t) + r_\ast(t) + r_0(t)]/r_{\rm frg}}$}, one obtains the expression for
the normal coordinate of the nebulous companion at time $t$, $z_0(t)$.

Adding the separation velocity to the list of free parameters makes their number one
too many, allowing no unique solution.  I show the dependence on the choice of the
heliocentric distance at separation in Table~4.  Comparison with Table~3 suggests
that the introduction of the separation velocity affects the magnitude of the
nongravitational acceleration only slightly, but shows that the magnitude of the
separation velocity is very sensitive to the choice of the heliocentric distance at
separation, increasing rapidly in both directions.  The acceleration --- both in the
radial and normal directions --- diminishes with increasing heliocentric distance at
separation when it exceeds 0.7~AU, but its moderate decrease is compensated by a
sharp increase in the radial separation velocity.  Over all, the introduction of the
separation velocity fails to make the solutions more realistic than before.  This
exercise also confirms that a separation velocity on the order of 10~m~s$^{-1}$ or
lower has practically no effect on the solution.

The last option in this category of solutions, based on the assumption that the
nebulous companion's motion relative to the main comet was triggered by the separation
velocity alone, with no contribution from the nongravitational acceleration, does
not help anything, as expected.  The radial motion is fitted best for a heliocentric
distance at separation of 0.77~AU (only about a week before Schmidt's October~10
observation), but the required radial separation velocity is fully 39~km~s$^{-1}$!

\section{Resolving the Problem}  
All attempts aimed at fitting Schmidt's (1882a) positional observations over three
consecutive days in terms of the motion of a single nebulous companion have failed
to offer a dynamically meaningful solution.  The feature described by Schmidt as
the {\small \bf western condensation} of the nebulous companion on October~10, 11,
and 12~UT must 
inevitably have been {\small \bf three different objects}.  Similarly, the eastern
condensation of the companion on October~11 could not possibly have been the same
mass as on October~12.  In addition to the inexplicably rapid motion relative to the
main comet, this conclusion is supported by at least two other pieces of evidence.
One, Schmidt's descriptions of the morphology varied from day to day, showing little
or no morphological similarity (Section~2.1).  And two, Barnard (1883b) suggested
that these features were omnipresent in the region southwest of the main comet in
mid-October, but complained about ``their visibility being so transient''
(Section~2.3).

I propose that the clue to solving the problem of nebulous companions is the key
role of {\small \bf terminal outburst}.  In general, outbursts, which last from
a fraction of a day to many days or weeks, are common manifestations of cometary
activity.  Some comets undergo giant explosions that appear to have no effect on
their well-being.  A quintessential example is a pair of huge outbursts, 40~days
apart, experienced by comet 41P/Tuttle-Giacobini-Kres\'ak in its 1973 return; in
either episode, the brightness soared by 8--9 magnitudes to the peak in a matter of
hours and then subsided at a similarly steep rate.\footnote{The light curve of comet
41P at its 1973 apparition is shown in Figure~7 of Sekanina (1984).}  Since 1973 the
comet has completed eight revolutions about the Sun apparently in perfect health.

By contrast, the expression {\it terminal outburst\/} derives from the observation
that this event often is experienced by cometary nuclei or their fragments shortly
before their {\it sudden disintegration\/}.  A possible mechanism for such a
cataclysmic event could be rotational instability triggered by torques from strongly
anisotropic outgassing, resulting eventually in rotational disruption, as discussed
by Jewitt (2021).  In line with the morphological diversity of the nebulous companions
of the Great September Comet, I hereby propose that they were triggered by {\small
\bf poorly-cemented minor fragments} of the comet's nucleus and seen only because
they happened to be {\small \bf caught in the brief terminal outburst}, when their
mass was suddenly shattered into {\small \bf clouds of} mostly {\small \bf microscopic
debris}.  Only a small fraction of nebulous companions was in fact detected not only
because of their transient nature, but also because only some minor fragments were
poorly cemented to the degree that they disintegrated this early after their birth;
other fragments could survive a greater part of the orbit or nearly the entire orbit.
   
The birth of minor fragments giving rise to the nebulous companions is linked to the
process of breakup of the comet's nucleus, which apparently occurred merely hours
after perihelion (Sekanina \& Chodas 2007).  Such minor fragments may have separated
in the aftermath of, rather than during, the prime event, an issue of no major import.
However, in order to end up degrees to the southwest of the comet's head, these
fragments must have been subjected to major nongravitational accelerations in both
the radial and normal directions.  The out-of-plane effect may have been the product
of high-obliquity rotation.  In any case, the {\small \bf similarity to the behavior
of the SOHO dwarf Kreutz sungrazers is striking}.  The only distinctions are their
different fragmentation histories and the SOHO sungrazers' absence of the terminal outburst,
unquestionably a consequence of extremely short lifetimes of both icy and dust grains
vigorously sublimating in the high-temperature environment of the solar corona.

The proposed outburst scenario implies a very short lifetime for the observed nebulous
companions, certainly less than 24~hours, which is why Schmidt's (1882a) observations
make sense only if they refer to unrelated objects.  Indeed, if a gradually sublimating
fragment is just of about the right size, its {\small \bf debris is detected only at
and near the peak of the terminal outburst} in the small telescopes used in 1882.  The
debris of fragments that are smaller than this critical size fails to reach the threshold
of detection even at the peak of the terminal outburst, thus remaining unobserved.  On
the other hand, for fragments of larger dimensions, which are fewer to begin with, the
process may take longer and be completed at larger heliocentric distances, thus escaping
detection.  Also, larger fragments are subjected to lower accelerations and they may not
venture as far away from the orbital plane as do smaller fragments.  In general, as
part of the process of nuclear fragmentation of the Great September Comet, the nebulous
companions appear to have represented a detached, transient extension to the diffuse
sheath of material that was observed to encompass the major nuclear fragments over
a protracted period of time. 

It is disappointing that none of the observers reported a magnitude estimate for
any of the nebulous companions.  Schmidt (1882a) noted that the feature he
observed appeared as a nebula ``easily perceptible in the finder'' of the
telescope,\footnote{The primary instrument of the Athens Observatory at the time
was a 16-cm f/11 Pl\"{o}ssl refractor, but in early September 1880 Schmidt
(1881a) acquired, on loan from the Academy of Sciences in Berlin, an 11-cm
f/13 Reinfelder \& Hertel refractor, which was installed on a patio in his Athens
residence.  The nebulous companions were observed with this telescope and its
finder.} but he offered no information on what the expression meant in
magnitude terms.  He even failed to specify the finder's aperture size.  Hartwig
(1883) observed the main comet with a 5-cm f/15 telescope and I presume that he
used this instrument for sighting the nebulous companion as well.  Markwick (1883)
employed a 7-cm achromatic for his Pietermaritzburg observations, while Barnard's
telescope was a 13-cm f/15 Byrne refractor.  I adopt a crude guess for an apparent
(as well as absolute)\footnote{The equivalence between the apparent and absolute
magnitudes is provided by the approximately balanced contributions to the brightness
from the heliocentric and geocentric distances.  At the time of Schmidt's October~10
observation the comet was, as Table~1 shows, at a heliocentric distance of 0.866~AU
and at a geocentric distance of 1.386~AU, which, given an $r^{-3.6}$ variation
(Sekanina 2002), leads to a difference of only 0.1~mag between the apparent and
absolute magnitudes.} magnitude~9 for the nebulous companions, inferred from
indirect information based on Schmidt's remarks on other objects he observed with
this finder.  The guess is in line with what he said about his first observation,
on 1880~December~30, of comet C/1880~Y1 (Pech\"{u}le).  Schmidt (1881b) wrote that
the comet was clearly seen in the finder of the 11-cm refractor, estimating its
condensation at magnitude~9.  Similarly, he remarked that comet C/1881~K1 (Tebbutt)
was an easy object in the same finder on 1881~October~13 (Schmidt 1882b); the
following day Pickering et al.\ (1900) measured the comet's brightness with a visual
photometer and found magnitude~9.2.  Schmidt (1881c) also felt comfortable with
using this same finder for examining stars of magnitude~9.  It appears that objects
of magnitude~9 were the finder's typical-brightness standard.  His observations of
C/1882~R1 also provided a constraint on the finder's limiting magnitude (Schmidt
1883); referring obviously to the comet's condensation, he noted that it was still
visible in the finder at the end of March 1883.  The light curve by Sekanina (2002)
suggests that its apparent magnitude at the time was 10.0, so that the limiting
magnitude of the finder may have been close to magnitude~11.

The estimated absolute magnitude of the nebulous companions allows one to provide
constraints on the mass consumed in the terminal outburst.  The peak visual
brightness of a disintegrating porous icy dust conglomerate depends on both the
released amount of molecular species that radiate in the visible spectrum, C$_2$
in particular, and on the total cross-sectional area of the disintegrated solid
material that scatters sunlight.  Because of a fairly small heliocentric distance,
atomic sodium could also contribute to the visual brightness but probably only
marginally.  Below I consider two extreme cases, the entire effect being due to
(i)~the abundance of C$_2$ molecules and (ii)~scattering by microscopic dust.

The first case employs A'Hearn \& Millis' (1980) relationships (i)~between the absolute
visual magnitude and the production rate of C$_2$ (normalized to 1~AU from the Sun)
that is required to sustain it; and (ii)~between the normalized production rates of
C$_2$ and H$_2$O, the latter approximated by the rate of OH.  Combined with the
standard water-ice sublimation model for rapidly rotating comets, one can use the
relationships to calculate the time needed to completely sublimate a disintegrated
fragment of a given size in an outburst with the peak absolute magnitude of 9 and
the corresponding amplitude of the outburst measured from a base level of brightness
given by outgassing from a spherical nucleus of the same dimensions.  For an assumed
bulk density of 0.5~g~cm$^{-3}$ I find that to reach the absolute magnitude~9, a
disintegrated spherical fragment 10~meters across needs near 1~AU from the Sun 1.0~hr
to sublimate away; 13~meters across needs 2.2~hr; 16~meters, 4.0~hr; and 20~meters,
7.9~hr.  The outburst amplitudes (relative to sunlight scattered by the surface of
the inactive object of the given dimensions) are, respectively, 12.5~mag, 11.8~mag,
11.2~mag, and 10.6~mag, so that the relevant absolute magnitudes vary, accordingly,
from 21.5~mag to 19.6~mag.  Because the photodissociation lifetime of C$_2$
at the pertinent heliocentric distances amounts to about 1~day or a little less
and because the observed lifetime of the nebulous companions is about the same (as
they are not seen for two days in a row), an outburst cannot last longer than a small
fraction of a day.  Correspondingly, the fragment's diameter at the outburst's onset
should not exceed $\sim$15~meters.\footnote{Other empirical formulas correlating the
water production rate with the magnitude (not necessarily the absolute magnitude;
e.g., Jorda et al.\ 1992) bypass the relationship with the C$_2$ production rate
and, by not clearly separating the contributions from C$_2$ and the dust, are less
appropriate.  On the other hand, A'Hearn \& Millis (1980) may have underestimated
the water production rate, in which case the upper limit should increase from
15~meters to about 20~meters across.}

In the other case a fragment that just lost all its volatiles is assumed to
disintegrate into an optically thin cloud of microscopic dust particles.  The
total projected cross-sectional area, which at a given particle albedo determines
the required brightness, is the sum of the individual particles' cross-sections.
At an assumed geometric albedo of 4~percent, the absolute magnitude~9 implies a
cross-sectional area of 8800~km$^2$.  If made up of dust grains 0.2~micron across
and density 3~g~cm$^{-3}$, the companion comes out to be $1.8 \times \! 10^9$\,g
in mass and, at a bulk density of 0.5~g~cm$^{-3}$, 24~meters initially across.
With the terminal dimensions now larger, this scenario is clearly inferior to the
C$_2$ based case.  Also, dust outbursts have a tendency to a slowly declining
post-peak brightness, inconsistent with the observational constraint.

The next issue is the rate at which the fragment was sublimating away between
separation and the terminal outburst.  A solution to this problem provides an estimate
of how much larger was the fragment at its birth than at the end of its lifespan.  The
mass{\vspace{-0.035cm}} sublimation rate, $\dot{Z}(t)$ (in g~cm$^{-2}$~day$^{-1}$),
makes the{\vspace{-0.065cm}} fragment's surface recede at a rate of $\dot{Z}(t)/\rho$
(in cm~day$^{-1}$), where $\rho$ is the bulk density.  The rate at which the fragment's
diameter, $D(t)$, is diminishing with time at $t$, is
\begin{equation}
dD(t) = -\frac{2\dot{Z}(t)}{\rho} dt.                                       
\end{equation}
I now consider the sublimation of water ice from a rapidly rotating spherical object,
in which case one can write approximately at the relevant heliocentric distances
\begin{equation}
\dot{Z}(t) = \dot{Z}_0 \left[ \frac{r_\oplus}{r(t)} \right]^2,            
\end{equation}
where \mbox{$r_\oplus = 1$ AU} and \mbox{$\dot{Z}_0 \simeq 0.8$ g\,cm$^{-2}$\,day$^{-1}$}.
Integrating from the time of separation, when the comet's distance from the Sun was
$r_{\rm frg}$ and the fragment's diameter $D_{\rm frg}$, to the time of terminal
outburst, when the distance was $r_{\rm fin}$ and the diameter $D_{\rm fin}$, one finds,
using Equation~(4),
\begin{eqnarray}
D_{\rm fin} & = & D_{\rm frg}-\frac{\dot{Z}_0 r_\oplus^2 \sqrt{2}}{k\rho}
 \!\int_{r_{\rm frg}}^{r_{\rm fin}} \!\!\!\frac{dr}{r \sqrt{r-q}} \nonumber \\
 & = & D_{\rm frg} - \frac{4 \dot{Z}_0 r_\oplus^2}{k \rho \sqrt{2q}} 
 \left( \!\arcsin \! \sqrt{\frac{q}{r_{\rm frg}}} \!-\! \arcsin \!\sqrt{\frac{q}{r_{\rm
 fin}}} \right)\!, \nonumber \\[-0.2cm]
 & &                                                                       
\end{eqnarray}
where $q$ is as before the comet's perihelion distance.  To find an absolute upper
limit on the effect, I put \mbox{$r_{\rm frg} \rightarrow q$} and \mbox{$r_{\rm
fin} < \infty$}.  With a reasonable value of the bulk density, \mbox{$\rho =
0.5$\,g\,cm$^{-3}$} one gets 
\begin{equation}
D_{\rm frg} < D_{\rm fin} + 47 \; {\rm meters}.                             
\end{equation}
Adopting, for example, \mbox{$D_{\rm fin} \simeq 18$ meters}, the initial diameter
of the fragment under the given conditions was less than 65~meters.  It is likely
that the effect was much smaller than this extreme estimate and so were the initial
fragment dimensions.  For example, with the observed \mbox{$r_{\rm fin} \simeq 0.9$
AU} and a modest \mbox{$r_{\rm frg} \simeq 0.1$ AU}, barely one day after perihelion,
the total amount by which the fragment's diameter contracted between the separation
and the terminal outburst was less than 6~meters, a small fraction of the final size.

The range of plausible heliocentric distances $r_{\rm frg}$ of fragments making up an
{\it observed\/} nebulous companion to the Great Comet can be constrained only partially.
I used Equations~(9) and (10) to compute, as a function of $r_{\rm frg}$, the magnitudes
of the radial and normal components of the nongravitational acceleration that fit the
positions of the western condensation of Schmidt's nebulous companion on, respectively,
October~10 and 12 as two unrelated objects, as shown in Table~5, and the position of
the eastern condensation on October~12, presented in Table~6.  The chosen acceleration
units, 10$^{-5}$\,AU day$^{-2}$ normalized to 1~AU from the Sun, are those employed
by Sekanina \& Kracht (2015) in their investigation of the SOHO Kreutz comets.  To
distinguish these absolute values from the dimensionless quantities $\Gamma_{\!\rm R}$
and $\Gamma_{\!\rm N}$ employed in Tables~3 and 4, I now use the designations
\begin{eqnarray}
\gamma_{\rm R} & = & \Gamma_{\!\rm R}(k/r_\oplus)^2 \nonumber \\[0.1cm]
\gamma_{\rm N} & = & \Gamma_{\!\rm N}(k/r_\oplus)^2.
\end{eqnarray}

\begin{table*}
\vspace{-4.15cm}
\hspace{0.55cm}
\centerline{
\scalebox{1}{
\includegraphics{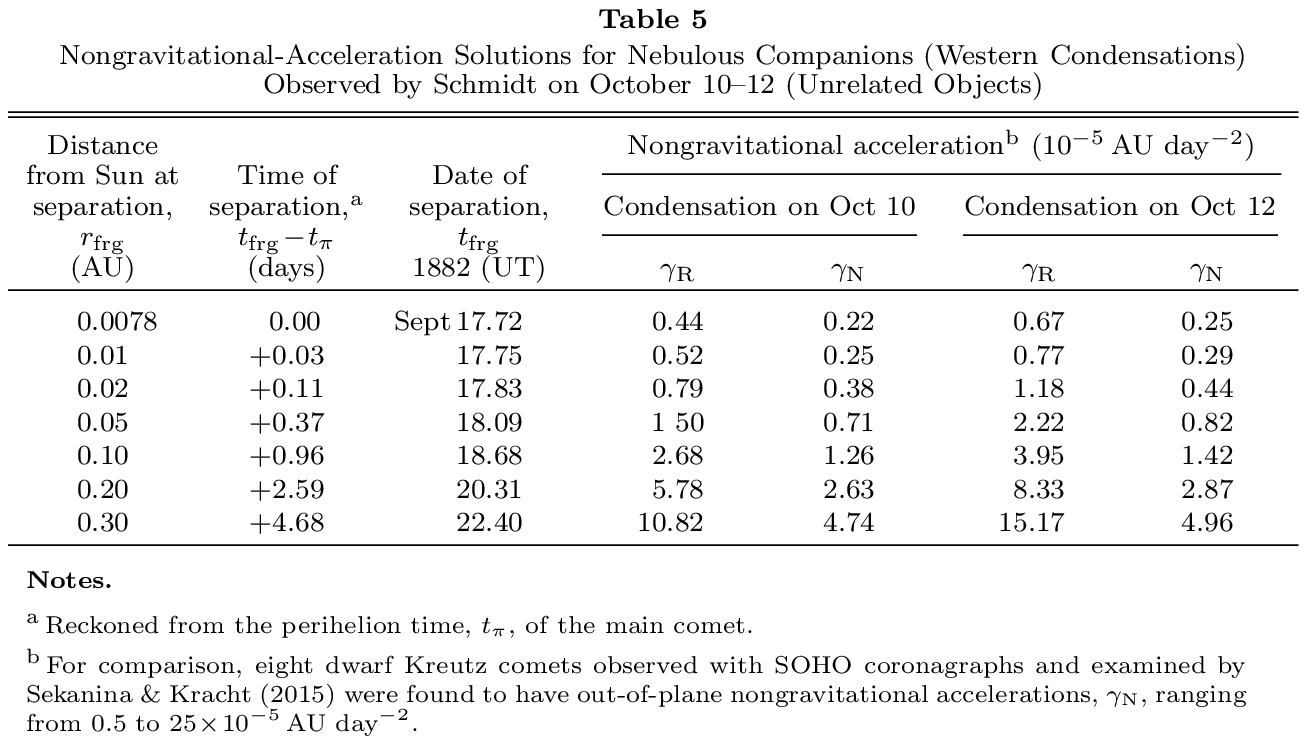}}}
\vspace{-17.6cm}
\end{table*}

It is noted that Equations~(9) and (10) were derived on the assumption that
\mbox{$r_{\rm frg} \gg q$}, which is not satisfied for the top few rows of Tables~5
and 6; the acceleration numbers thus become increasingly inaccurate as $r_{\rm frg}$
decreases, most burdened with errors being the numbers for \mbox{$r_{\rm frg} = q$}.
Numerical integrations suggest that for a separation at perihelion, the correct
accelerations are about $\frac{2}{3}$ the tabulated values.

Two effects influencing the nongravitational acceleration are unaccounted for.  The
acceleration is assumed to have been constant for a given separation time, but it
was actually increasing as the fragment's dimensions were diminishing with time.  This
effect was of course minor if the fragment lost little mass between the separation
and the terminal outburst.  The other effect is the fragment's probable progressive
crumbling during flight, which also gradually increases the effective value of the
nongravitational acceleration, a complex problem that is not addressed here in any
detail.

While it is virtually certain that the fragments triggering the observed nebulous
companions broke off from the comet's nucleus after perihelion (and after its
primary splitting event), only soft constraints are possible about their separation
times.  It is unlikely that the fragments were released farther from the Sun than
about 0.3~AU, because their nongravitational accelerations would then be too high,
comparable with the Sun's gravitational acceleration, for objects in the mass range
estimated above from their brightness.  On the other hand, the fragments probably
were not released in the immediate proximity of perihelion, because their
nongravitational accelerations would suggest objects too massive to be as abundant
and transient as their observations appear to suggest.  The plausible intermediate
heliocentric distances at separation, crudely between, say, 0.05~AU and 0.3~AU,
imply radial and out-of-plane nongravitational accelerations of a few units of
10$^{-5}$\,AU~day$^{-2}$ at 1~AU from the Sun, about one order of magnitude lower
than the Sun's gravitational acceleration and typical for the dwarf Kreutz sungrazers,
seen in large numbers in the images taken with the coronagraphs onboard the SOHO
space probe.

Finally, there is the issue of the sign of the out-of-plane component of the
nongravitational acceleration that the fragments triggering the nebulous companions
are subjected to.  Given the retrograde sense of the orbital motion of the Great
September Comet, the companions' locations to the southwest of the comet's head
imply the general direction to the comet's northern orbital pole, so that
\mbox{$\gamma_{\rm N} > 0$}.  Inspection of the select set of 193 SOHO Kreutz
sungrazers with quality orbits (Sekanina 2021) suggests that 70 percent of the
members of Population~II, to which the Great September Comet belongs, were
subjected to negative $\gamma_{\rm N}$ accelerations.  This is not necessarily
an indication of disparity between the fragments of the Great September Comet
and the SOHO sungrazers, which are surviving products of fragmentation in the
previous revolution about the Sun.  If the direction of the nongravitational
acceleration is rotation-related, the preperihelion (SOHO) and post-perihelion
(nebulous companions) fragments may be expected to be released in opposite
out-of-orbit directions for certain spin-axis orientations.  However, regardless
of the acceleration mechanism, there should exist nebulous companions located,
approximately symmetrically, to the northwest of the comet's head.  The fact
that none was reported is probably a statistical quirk, given the small number
of observed cases.

\begin{table}[b]
\vspace{-3.5cm}
\hspace{5.25cm}
\centerline{
\scalebox{1}{
\includegraphics{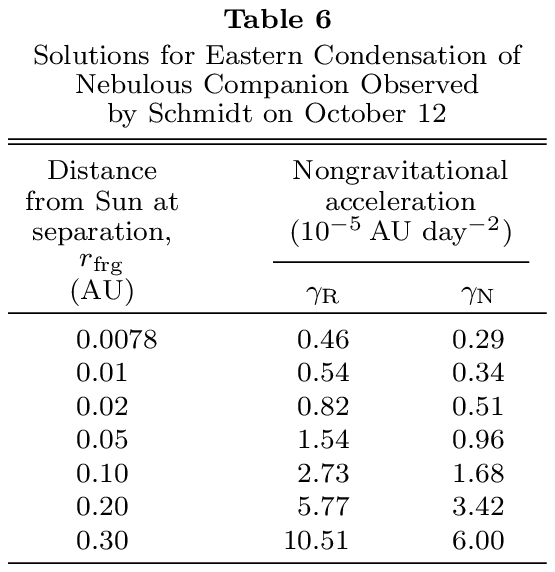}}}
\vspace{-19.8cm}
\end{table}

\section{Conclusions}  
Large numbers of minor fragments of the main comet's nucleus, byproducts of both
the primary breakup shortly after perihelion and the subsequent cascading crumbling
of the secondary nuclei, are presumed to have remained undetected because of their
intrinsic faintness.  Exceptions were decameter-sized fragments that happened to
be caught in terminal outburst, signaling their sudden disintegration perhaps
on account of rotational disruption by outgassing torques.  The resulting clouds
of debris, typically several arcminutes in diameter, located degrees to the
southwest of the comet's head and estimated here at having reached, on the
average, magnitude~9 during the proposed outburst's brief peak, were reported
as transient nebulous companions independently by four observers between 1882
October~5 and 14, less than one month after perihelion.  To end up that far from
the comet's nucleus, the fragments' motions must have been subjected to a major
outgassing-driven nongravitational acceleration with a significant out-of-plane
component.  I assume the sublimation of water ice, but admixtures of other ices
would not fundamentally change the outcome.  The proposed scenario implies that
each event could be under observation with a small telescope over a period of
time of less than one day; Schmidt's (1882a) report of a nebulous companion seen
in consecutive mornings must have referred to unrelated objects.

Predicated on the strength of the positional data and estimated photometry on the
nebulous companions are the conclusions that the minor fragments, tens of meters
across, had the brightness probably dominated by C$_2$ molecular emissions and that
their nongravitational accelerations with the significant out-of-plane component were
on the order of 10$^{-5}$\,AU~day$^{-2}$ when normalized to 1~AU from the Sun.  The
fragment dimensions and accelerations are comparable to those for the Kreutz dwarf
sungrazers observed with the SOHO's coronagraphs.  The only differences between the
Great September Comet's fragments triggering the nebulous companions and the SOHO
Kreutz sungrazers moments before their preperihelion disappearance are that (i)~they
do not share the same fragmentation history and (ii)~the latter experience no terminal
outburst because of extremely short lifetimes of rapidly sublimating icy and dust
grains in the Sun's corona.  Even though it may come as a surprise, it appears that
objects bearing a strong resemblance to the dwarf Kreutz sungrazers from the SOHO's
coronagraphic images were, in their final stage of disintegration, repeatedly detected
with small-aperture telescopes more than a century before the launch of the SOHO
mission.\\

This research was carried out at the Jet Propulsion Laboratory, California Institute of
Technology, under contract with the National Aeronautics and Space Administration.\\[-0.1cm]
\begin{center}
{\footnotesize REFERENCES} \\[0.1cm]
\end{center}
\begin{description}
{\footnotesize
\item[\hspace{-0.3cm}]
A'Hearn, M. F., \& Millis, R. L. 1980, AJ, 85, 1528
\\[-0.57cm]
\item[\hspace{-0.3cm}]
Barnard, E. E. 1882, Sid. Mes., 1, 221
\\[-0.57cm]
\item[\hspace{-0.3cm}]
Barnard, E. E. 1883a, Astron. Nachr., 104, 267
\\[-0.57cm]
\item[\hspace{-0.3cm}]
Barnard, E. E. 1883b, Sid. Mes., 1, 255
\\[-0.57cm]
\item[\hspace{-0.3cm}]
Boehnhardt, H. 2004, in Comets II, ed. M. Festou, H. U. Keller, \&{\linebreak}
 {\hspace*{-0.6cm}}H. A. Weaver (Tucson, AZ: University of Arizona), 301
\\[-0.57cm]
\item[\hspace{-0.3cm}]
Brooks, W. R. 1883, Sid. Mes., 2, 149
\\[-0.57cm]
\item[\hspace{-0.3cm}] 
Frost, E. B. 1927, Mem. Nat. Acad. Sci., 21, No.\,14
\\[-0.57cm]
\item[\hspace{-0.3cm}]
Hartwig, E. 1883, Astron. Nachr., 106, 225
\\[-0.57cm]
\item[\hspace{-0.3cm}]
Hind, J. R. 1883, Sid. Mes., 1, 259
\\[-0.57cm]
%
%
\item[\hspace{-0.3cm}]
Jewitt, D. 2021, AJ, 161, 261
\\[-0.57cm]
\item[\hspace{-0.3cm}]
Jorda, L., Crovisier, J., \& Green, D. W. E. 1992, in Asteroids,{\linebreak}
 {\hspace*{-0.6cm}}Comets, Meteors 1991, ed. A. W. Harris \& E. Bowell (Houston,{\linebreak}
 {\hspace*{-0.6cm}}TX: Lunar and Planetary Institute), 285
\\[-0.57cm]
\item[\hspace{-0.3cm}]
Kreutz, H. 1882, Astron. Nachr., 103, 209
\\[-0.57cm]
\item[\hspace{-0.3cm}]
Markwick, E. E. 1883, MNRAS, 43, 322
\\[-0.57cm]
%
%
\item[\hspace{-0.3cm}]
Oppenheim, H. 1882, Astron. Nachr., 103, 283
\\[-0.57cm]
\item[\hspace{-0.3cm}]
Payne, W. W. 1883, Sid. Mess., 2, 192
\\[-0.57cm]
\item[\hspace{-0.3cm}]
Pickering, E. C., Searle, A., \& Wendell, O. C. 1900, Ann. Harv.{\linebreak}
 {\hspace*{-0.6cm}}Coll. Obs., 33, 149
\\[-0.57cm]
\item[\hspace{-0.3cm}]
Schmidt, J. F. J. 1881a, Astron. Nachr., 99, 101
\\[-0.57cm]
\item[\hspace{-0.3cm}]
Schmidt, J. F. J. 1881b, Astron. Nachr., 99, 253
\\[-0.57cm]
\item[\hspace{-0.3cm}]
Schmidt, J. F. J. 1881c, Astron. Nachr., 99, 349
\\[-0.57cm]
\item[\hspace{-0.3cm}]
Schmidt, J. F. J. 1882a, Astron. Nachr., 103, 305
\\[-0.57cm]
\item[\hspace{-0.3cm}]
Schmidt, J. F. J. 1882b, Astron. Nachr., 101, 249
\\[-0.37cm]
\item[\hspace{-0.3cm}]
Schmidt, J. F. J. 1883, Astron. Nachr., 105, 341
\\[-0.57cm]
\item[\hspace{-0.3cm}]
Sekanina, Z. 1982, in Comets, ed. L. L. Wilkening (Tucson, AZ:{\linebreak}
 {\hspace*{-0.6cm}}University of Arizona), 251
\\[-0.57cm]
\item[\hspace{-0.3cm}]
Sekanina, Z. 1984, Icarus, 58, 81
\\[-0.57cm]
\item[\hspace{-0.3cm}]
Sekanina, Z. 2002, ApJ, 566, 577
\\[-0.57cm]
\item[\hspace{-0.3cm}]
Sekanina, Z. 2021, arXiv eprint 2109.01297
\\[-0.57cm]`
\item[\hspace{-0.3cm}]
Sekanina, Z., \& Chodas, P. W. 2007, ApJ, 663, 657
\\[-0.57cm]
%
%
\item[\hspace{-0.3cm}]
Sekanina, Z., \& Kracht, R. 2015, ApJ, 801, 135
\\[-0.64cm]
%
%
\item[\hspace{-0.3cm}]
Zelbr, K. 1882, Sitz. Akad. Wiss. Wien, 86, 1090}
\end{description}
%
%
\end{document}